\documentclass[journal = cmatex]{achemso}
\setkeys{acs}{articletitle = true}
\setkeys{acs}{maxauthors = 100}
\usepackage[T1]{fontenc}
\usepackage{lineno}

\author{Shanyuan Niu}
\author{Boyang Zhao}
\affiliation{Mork Family Department of Chemical Engineering and Materials Science, University of Southern California, Los Angeles, CA, USA}
\author{Kevin Ye}
\affiliation{Department of Materials Science and Engineering, Massachusetts Institute of Technology, Cambridge, MA, USA}
\author{Elisabeth Bianco}
\affiliation{Materials and Manufacturing Directorate, Air Force Research Laboratory, Wright-Patterson AFB, Dayton, OH, USA.}
\author{Jieyang Zhou}
\affiliation{Mork Family Department of Chemical Engineering and Materials Science, University of Southern California, Los Angeles, CA, USA}
\author{Michael E. McConney}
\affiliation{Materials and Manufacturing Directorate, Air Force Research Laboratory, Wright-Patterson AFB, Dayton, OH, USA.}
\author{Charles Settens}
\affiliation{Materials Research Laboratory, Massachusetts Institute of Technology, Cambridge, MA, USA}
\author{Ralf Haiges}
\affiliation{Loker Hydrocarbon Research Institute and Department of Chemistry, University of Southern California, Los Angeles, California, USA}
\author{R. Jaramillo}
\affiliation{Department of Materials Science and Engineering, Massachusetts Institute of Technology, Cambridge, MA, USA}
\author{Jayakanth Ravichandran}
\email{jayakanr@usc.edu}
\affiliation{Mork Family Department of Chemical Engineering and Materials Science, University of Southern California, Los Angeles, CA, USA}
\alsoaffiliation{Ming Hsieh Department of Electrical Engineering, University of Southern California, Los Angeles, CA, USA}

\title{Crystal growth and structural analysis of perovskite chalcogenide BaZrS$_3$ and Ruddlesden-Popper phase Ba$_3$Zr$_2$S$_7$}

\keywords{crystal growth, crystallographic structure, flux growth}

\begin{document}

\begin{abstract}
Perovskite chalcogenides are gaining substantial interest as an emerging class of semiconductors for optoelectronic applications. High quality samples are of vital importance to examine their inherent physical properties. We report the successful crystal growth of the model system, BaZrS$_3$ and its Ruddlesden-Popper phase Ba$_3$Zr$_2$S$_7$ by flux method. X-ray diffraction analyses showed space group of $Pnma$ with lattice constants of $a$ = 7.056(3) \AA\/, $b$ = 9.962(4) \AA\/, $c$ = 6.996(3) \AA\/ for BaZrS$_3$ and $P4_2/mnm$ with $a$ = 7.071(2) \AA\/, $b$ = 7.071(2) \AA\/, $c$ = 25.418(5) \AA\/ for Ba$_3$Zr$_2$S$_7$. Rocking curves with full-width-at-half-maximum of 0.011$^\circ$ for BaZrS$_3$ and 0.027$^\circ$ for Ba$_3$Zr$_2$S$_7$ were observed. Pole figure analysis, scanning transmission electron microscopy images and electron diffraction patterns also establish high quality of grown crystals. The octahedra tilting in the corner-sharing octahedra network are analyzed by extracting the torsion angles.


\end{abstract}


\section{Introduction}
Perovskite chalcogenides have been gaining increased attention as a class of emerging semiconductors with rich tunability and excellent optoelectronic properties in the visible to infrared spectrum.\cite{Sun:2015be,Wang:2016fh,Kuhar:2017gw,Ju:2017iw,Niu:2018doa,Niu:2018hu,Filippone:2018cr,Niu:2018cx,Hanzawa:2019cf,Swarnkar:2019ia} Perovskite chalcogenides can be viewed as the inorganic alternatives to hybrid organic-inorganic halide perovskites, with stable, benign, abundant composition, and ultrahigh absorption coefficients. On the other hand, perovskite chalcogenides can also be viewed as the chalcogenide counterparts of perovskite oxides, with much lower bandgap and improved response to visible and infrared light.\cite{Bennett:2009hn,Sun:2015be} The combination of ultrahigh absorption coefficient, good carrier mobility, along with tunable band gap, good thermal and aqueous stability, and benign, earth abundant composition creates opportunities for a broad range of photonic, optoelectronic, and energy applications, including solar cells,\cite{Sun:2015be,Wang:2016fh,Ju:2017iw,Niu:2018hu} photodetectors,\cite{Niu:2018doa,Niu:2018cx,Swarnkar:2019ia} lighting devices,\cite{Hanzawa:2019cf} and photoelectrochemical devices.\cite{Kuhar:2017gw}

These materials were known to exist for a long time and the synthetic efforts can date back to over half a century ago.\cite{Hahn:1956wo,CLEARFIELD:1963ip,LELIEVELD:1980wp,Huster:1980fn} Synthetic methods of ceramic samples include heating binary sulfide mixture for several weeks,\cite{Lee:2005et} sulfurization of corresponding oxides with CS$_2$\cite{BinOkai:1988ut,CLEARFIELD:1963ip} or H$_2$S.\cite{LELIEVELD:1980wp}, and solid state reaction with catalytic addition of iodine.\cite{niu:2016ce} Crystal growth efforts have been reported for perovskite related Ruddlesden-Popper phases\cite{Hung:1997gg,Chen:1994ba,Chen:1993cy} and hexagonal phases.\cite{Huster:1980fn} However, most reports on perovskite chalcogenides were limited to the structural studies of polycrystalline samples. Recently, these materials were rediscovered and explored as a class of semiconductors for optoelectronic applications. Experimental explorations of relevant physical properties started on the prototypical perovskite chalcogenide, BaZrS$_3$.\cite{Meng:2016dv,Perera:2016gf,niu:2016ce,Gross:2017bo,Niu:2018hr} Synthesis of high quality samples are of vital importance in such explorations. \citet{Meng:2016dv} synthesized BaZrS$_3$ by conventional solid-state reaction of binary mixtures with repeated annealing. \citet{Perera:2016gf} synthesized BaZrS$_3$ by high temperature sulfurization of oxides with CS$_2$. We have demonstrated synthesis of BaZrS$_3$ with catalytic addition of iodine in solid state reaction to enable one-shot, shorter synthesis.\cite{niu:2016ce} However, growth of single crystals for BaZrS$_3$ is yet to be demonstrated, which is necessary to study the intrinsic optical properties of this model perovskite chalcogenide material. In this work, we will report the crystal growth of BaZrS$_3$, along with its $n=2$ Ruddlesden-Popper phase Ba$_3$Zr$_2$S$_7$.

\section{Results and discussion}
\subsection{Crystal growth and morphology}
\begin{figure}[ht]
\centering
\includegraphics[width=5.5 in]{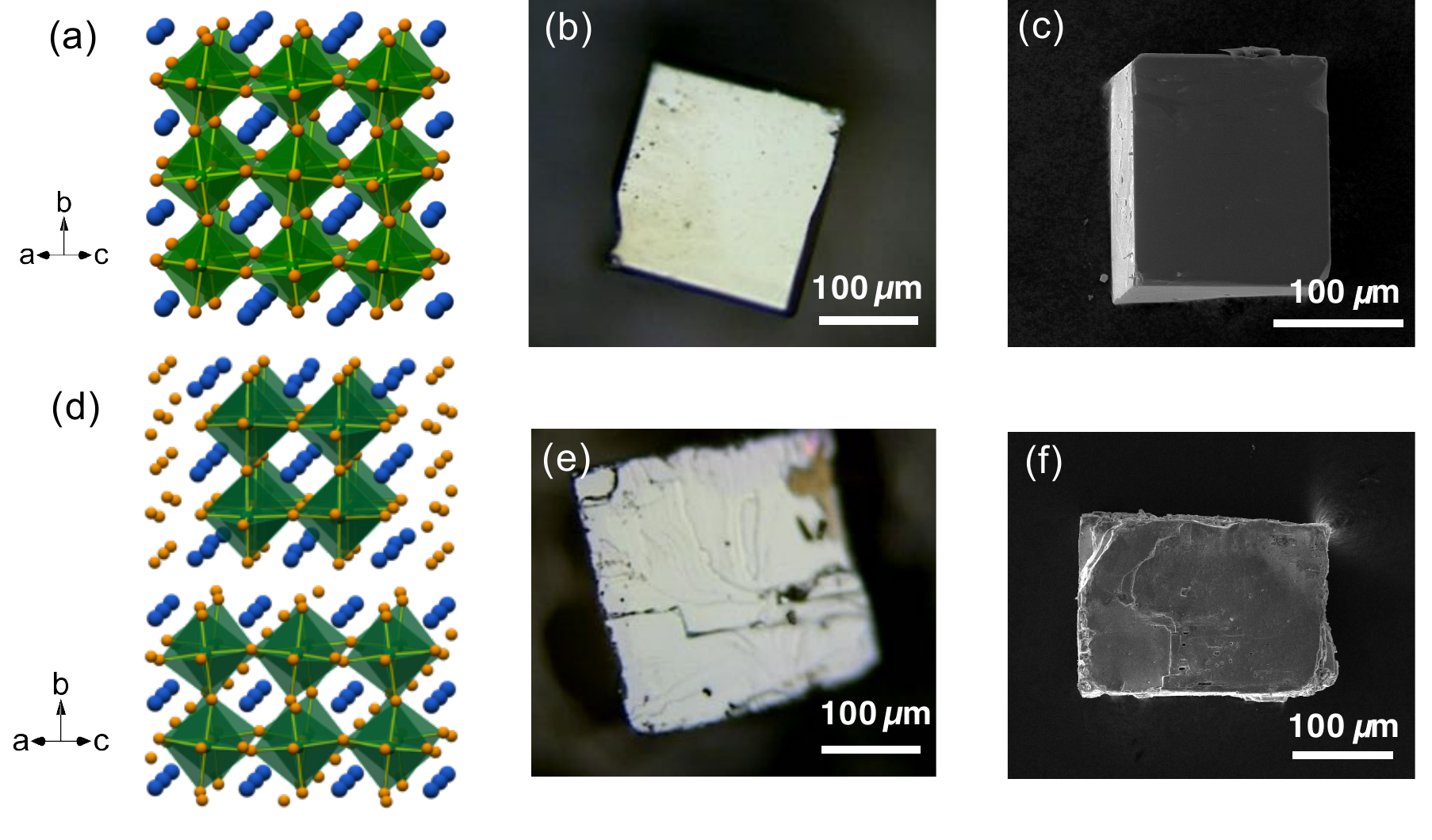}
\caption{\label{Fig:Fig1} Schematic crystal structure for (a) BaZrS$_3$ and (d) Ba$_3$Zr$_2$S$_7$. The blue, orange, and green spheres represent Ba, S, and Zr, respectively. The ZrS$_6$ octahedra are highlighted. Optical pictures of (b) BaZrS$_3$, (e) Ba$_3$Zr$_2$S$_7$ crystals. SEM images of (c) BaZrS$_3$, (f) Ba$_3$Zr$_2$S$_7$ crystals.}
\end{figure}

The single crystals were grown using salt flux method with BaCl$_2$ flux. We have also tried growing these crystals with other salts, such as KI and BaCl$_2$/MgCl$_2$ eutectic. The results were similar, the salt removing process was notably easier, but the obtained crystals were not as large. In the case of BaZrS$_3$, the predominant morphology of the obtained crystals was cube-like with sharp edges and well-defined surfaces that correspond to crystal facets, as shown in the images in Fig.\ref{Fig:Fig1}(b),(e). The obtained BaZrS$_3$ crystals were up to 500 $\mu$m in size, and showed shiny/reflective metallic luster when looked at normal incidence. In the case of RP phases, the attempt for single phase crystal growth by varying the stoichiometry of precursors was not successful. Salt flux growth is a slow process, and the energy barriers to get across RP phase boundaries are presumably shallow as they possess similar chemistry and structure. As a result, obtained crystallites were a mixture of RP phase, with the relative yield affected by the starting stoichiometry. For Ba$_3$Zr$_2$S$_7$, the predominant crystal morphology was square platelets, with similar shiny metallic luster on the surface. However, when looked under microscope, the top platelet crystal facets are not as well defined, and show layered like terraces on the surfaces, as shown in Fig.\ref{Fig:Fig1}(c),(f), presumably due to the stacking of perovskite slabs and rock-salt layer along the $c$ axis. The isolated Ba$_3$Zr$_2$S$_7$ crystal pieces were up to 300 um in size. A small amount of Ba$_4$Zr$_3$S$_{10}$ crystallites were also found in certain cases. However, the effort to consistently grow large Ba$_4$Zr$_3$S$_{10}$ crystals was not successful.

\subsection{Crystal structure of BaZrS$_3$}
We have performed single crystal X-ray diffraction analysis for the crystals. For BaZrS$_3$, we observed twinning in the BaZrS$_3$ crystals and had to break the crystals into smaller pieces for a reasonable solution to the diffraction analysis. A lustrous dark red plate-like specimen of BaZrS$_3$, approximate dimensions 0.005 mm $\times$ 0.008 mm $\times$ 0.008 mm, was used for the X-ray crystallographic analysis. The measurement was performed at 100 K. The integration of the data using an orthorhombic unit cell yielded a total of 4218 reflections to a maximum $\theta$ angle of 30.54$^\circ$ (0.70 \AA\/ resolution), of which 798 were independent (average redundancy 5.286, completeness = 94.8\%, R$_{int}$ = 5.33\%, R$_{sig}$ = 4.55\%) and 623 (78.07\%) were greater than 2$\sigma$($F^2$). The final cell constants of $a$ = 7.056(3) \AA\/, $b$ = 9.962(4) \AA\/, $c$ = 6.996(3) \AA\/, volume = 491.8(3) \AA\/$^3$, are based upon the refinement of the XYZ-centroids of 1919 reflections above 20$\sigma$($I$) with 9.169$^\circ$ $<$ 2$\theta$ $<$ 60.88$^\circ$. Data were corrected for absorption effects using the multi-scan method (SADABS). The ratio of minimum to maximum apparent transmission was 0.764. The calculated minimum and maximum transmission coefficients (based on crystal size) are 0.7310 and 0.8970. The structure was solved and refined using the space group $Pnma$, with Z = 4 for the formula unit, BaZrS$_3$. The final anisotropic full-matrix least-squares refinement on $F^2$ with 29 variables converged at $R_1$ = 3.36\%, for the observed data and $wR_2$ = 7.57\% for all data. The goodness-of-fit was 1.162. The largest peak in the final difference electron density synthesis was 1.804 e$^-$/\AA\/$^3$ and the largest hole was -1.788 e$^-$/\AA\/$^3$ with an root-mean-square deviation of 0.356 e$^-$/\AA\/$^3$. On the basis of the final model, the calculated density was 4.386 g/cm$^3$ and F(000), 576 e$^-$.

\begin{table}[ht]
\scriptsize
\begin{center}
\begin{tabular*}{\columnwidth}{ @{\extracolsep{\fill}\hspace{\tabcolsep}} lllllll @{\hspace{\tabcolsep}}}
\hline 
\hline
\multicolumn{2}{l}{Space Group} & $Pnma$ &  \multicolumn{2}{r}{Temperature} & 100 K \\ 
\hline 
\multicolumn{2}{l}{Lattice} &  $a$(\AA\/) & $b$(\AA\/) & $c$(\AA\/) \\  
\multicolumn{2}{l}{Constants} &  7.056(3) & 9.962(4) & 6.996(3) \\ 
\hline 
\multicolumn{2}{l}{Atomic Sites} & $x/a$ & $y/b$ & $z/c$ & U(eq) \\ 
\multicolumn{2}{l}{Ba1} & 0.54544(10) & 0.25 & 0.99030(10) & 0.00696(17) \\
\multicolumn{2}{l}{S1} & 0.4949(4) & 0.25 & 0.4382(4) & 0.0070(6) \\ 
\multicolumn{2}{l}{S2} & 0.2896(3) & 0.46782(19) & 0.7896(3) & 0.0060(4) \\
\multicolumn{2}{l}{Zr1} & 0.5 & 0.5 & 0.5 & 0.0031(2) \\ 
\hline
& U$_{11}$  & U$_{22}$ & U$_{33}$ & U$_{23}$ & U$_{13}$ & U$_{12}$ \\
\hline
Ba1 & 0.0091(3) & 0.0040(3) & 0.0078(3) & 0 & -0.0004(3) & 0 \\
S1 & 0.0101(14) & 0.0029(12) & 0.0079(13) & 0 & 0.0008(9) & 0 \\
S2 & 0.0075(9) & 0.0054(8) & 0.0052(9) & 0.0012(7) & 0.0015(7) & -0.0009(8) \\
Zr1 & 0.0038(5) & 0.0021(4) & 0.0035(5) & 0.0001(4) & 0.0001(5) & 0.0000(3) \\
\hline 
\hline 
\end{tabular*} 
\caption{Space group, lattice constants, atomic coordinates, equivalent isotropic atomic displacement parameters (\AA\/$^2$), and anisotropic atomic displacement parameters (\AA\/$^2$) from X-ray crystallographic analysis for BaZrS$_3$.}
\label{Tab:BZSXtalXRD}
\end{center}
\end{table}

The detailed structural parameters are shown in Table \ref{Tab:BZSXtalXRD}. Based on these parameters, schematic crystal structures were constructed, as shown in Fig.\ref{Fig:Fig1}(a). BaZrS$_3$ is isostructural with GdFeO$_3$ and adopts the distorted perovskite structure with a space group of $Pnma$, which agrees well with previous polycrystalline studies.\cite{CLEARFIELD:1963ip,LELIEVELD:1980wp,niu:2016ce} The octahedra share corners to form a three-dimensional network. This structure differs from ideal cubic perovskite in several ways. If we consider a ``pseudo-cubic'' structure with three axes pointing along the directions the octahedra are connecting corners, while the $b$ axis of the orthorhombic cell is aligned with one of the directions, $a$ and $c$ axes are close to the face diagonals of the other two pseudo-cubic axes. The pseudo-cubic cell constant, or effectively the spacing between octahedra, is 4.968\AA\/ along [101] and [10$\overline{1}$] directions, and 4.981\AA\/ spacing along [010] direction in BaZrS$_3$. Thus, to grow high quality epitaxial films of BaZrS$_3$, one can use substrates with a square symmetry and lattice constants of around 4.97\AA\/ for commensurate epitaxy or around 3.51\AA\/ for incommensurate epitaxy.

\begin{figure}[ht]
\centering
\includegraphics[width=4.5 in]{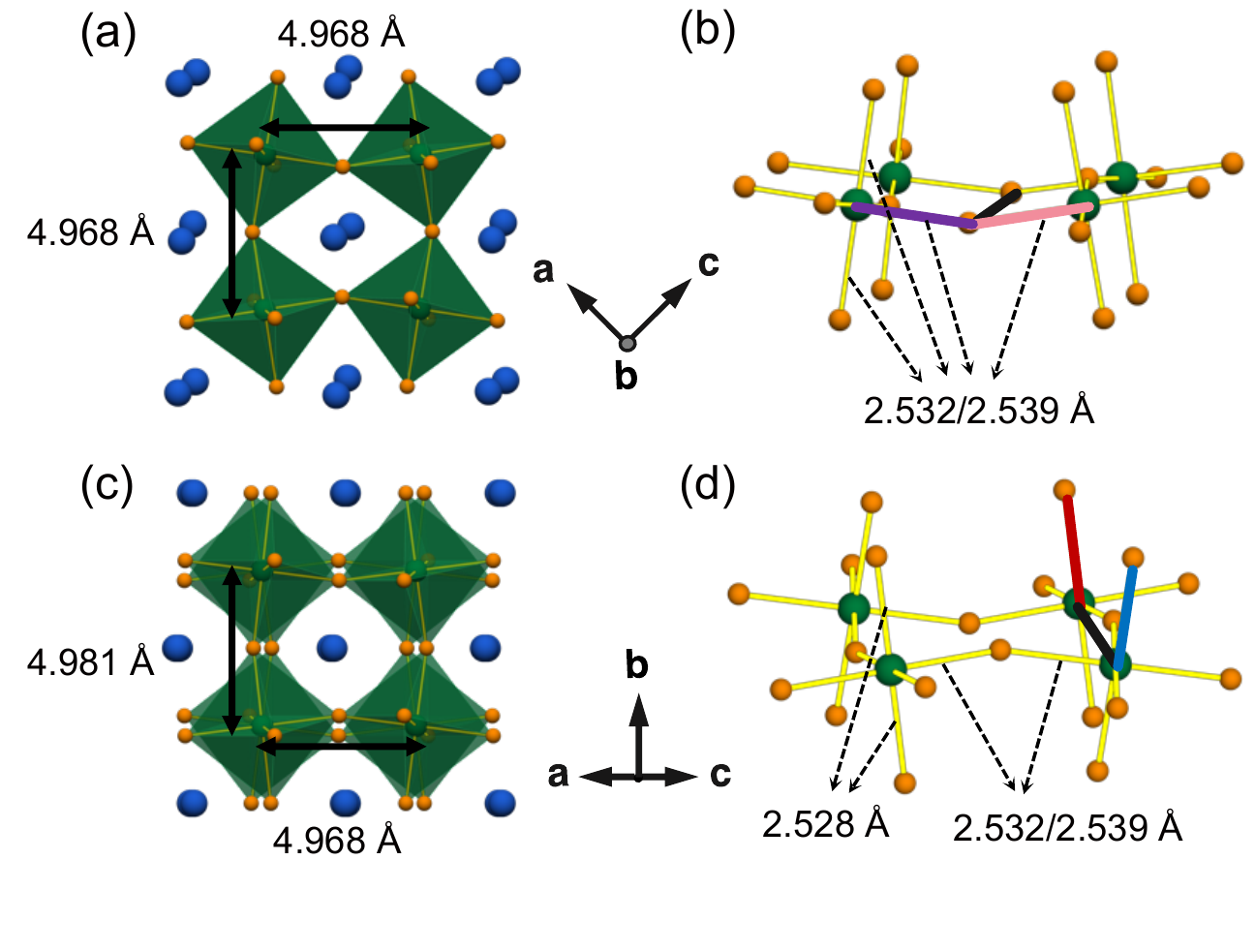}
\caption{\label{Fig:Tilt} (a) Schematic structure viewed along $b$ axis in BaZrS$_3$ shows the in phase tilt where the tilting of adjacent octahedra layers are overlapping. (b) Ball-stick model in perspective view showing the Zr-S-S-Zr torsion angle (purple-black-pink) to calculate the in-phase tilting angle. (c) Schematic structure viewed along (101) direction in BaZrS$_3$ shows the out of phase tilt where the tilting of adjacent layers are in opposite direction. (d) Ball-stick model in perspective view showing the S-Zr-Zr-S torsion angle (blue-black-red) to calculate out-of-phase tilting angle. The octahedra spacing (Zr-Zr distance) and octahedra shape (Zr-S bond length) along different directions are marked in the schematics.}
\end{figure}

\subsection{Octahedra tilting in BaZrS$_3$}
One key feature in BaZrS$_3$ crystal structure is the tilting of ZrS$_6$ octahedra. The tilting along $b$ axis is in phase, with adjacent octahedra layers in $ac$ plane rotating towards the same direction and remain perfectly overlapped when viewed along $b$ axis, as shown in Fig.\ref{Fig:Tilt}(a). The tilting along [101] and [10$\overline{1}$] directions are out of phase, with adjacent octahedra layers twisted in different directions, as shown in Fig.\ref{Fig:Tilt}(c). The tilting amplitude can be calculated by analysing the torsion angle of the Zr-S bonds. For in-phase tilt, the torsion angle is half of difference between 180$^\circ$ and the Zr-S-S-Zr torsion angle linking octahedra in the same layer, as shown in Fig.\ref{Fig:Tilt}(c). For out-of-phase tilt, the torsion angle is half of the S-Zr-Zr-S torsion angle linking octahedra in two adjacent layers, as shown in Fig.\ref{Fig:Tilt}(d). The extracted tilting angle along [101] and [10$\overline{1}$] directions are the same, -7.027$^\circ$ (negative denoting out-of-phase tilt) and while the tilting angle along [010] is 9.0005$^\circ$ (positive denoting in-phase tilt). The obtained results show that BaZrS$_3$ adopts a Glazer tilting system of $a^-b^+a^-$ (or standardized $a^+b^-b^-$).\cite{Woodward:1997ci} It is worth noting that although Zr is sitting in the center of ZrS$_6$ octahedra, the octahedra are slightly distorted, with three pairs of Zr-S bonds of bond length 2.532\AA\/, 2.539\AA\/, and 2.528\AA\/, respectively. The shortest pair of bonds are along [010], while octahedra with pairs of 2.532\AA\/ and 2.539\AA\/ Zr-S bonds are alternated along [101] and [10$\overline{1}$]. If we compare [010] to [101]/[10$\overline{1}$], the shorter Zr-S bonds (2.528\AA\/ compared to 2.532\AA\/ and 2.539\AA\/) and larger octahedra (Zr-Zr) spacing (4.981\AA\/ compared to 4.968\AA\/) is enabled by a smaller Zr-S-Zr bond angle (160.236$^\circ$ compared to 156.905$^\circ$), due to the larger octahedra tilting amplitude around [010] (9.0005$^\circ$ compared to 7.027$^\circ$). The in-phase tilting around $b$ axis also results in net displacement of S atoms, and consequently displacement of Ba atoms in $ac$ plane.

\subsection{Crystal structure of Ba$_3$Zr$_2$S$_7$}
For Ba$_3$Zr$_2$S$_7$, previous studies identified several possible structural variants, $I4/mmm$\cite{Chen:1994ba}, $P4_2/mnm$\cite{Hung:1997gg}, and $Cccm$\cite{Saeki:1991fa}. We performed single-crystal XRD studies on the grown crystals at 100 K and found that the crystals adopted the $P4_2/mnm$ space group, as shown in Table \ref{Tab:BZS327XtalXRD}. A dark red prism-like piece of Ba$_3$Zr$_2$S$_7$ crystal, approximate dimensions of 0.056 mm $\times$ 0.086 mm $\times$ 0.118 mm, was used for the X-ray crystallographic analysis in the same setup. The integration of the data using a tetragonal unit cell yielded a total of 29151 reflections to a maximum $\theta$ angle of 28.32$^\circ$ (0.75 \AA\/ resolution), of which 898 were independent (average redundancy 28.762, completeness = 99.7\%, R$_{int}$ = 4.3\%, R$_{sig}$ = 1.14\%) and 743 (82.74\%) were greater than 2$\sigma$($F^2$). The structure was solved and refined using the space group $P4_2/mnm$, with Z = 4 for the formula unit, Ba$_3$Zr$_2$S$_7$. The final cell constants of $a$ = 7.071(2) \AA\/, $b$ = 7.071(2) \AA\/, $c$ = 25.418(5) \AA\/, $\alpha$ = 90$^\circ$, $\beta$ = 90$^\circ$, $\gamma$ = 90$^\circ$, volume = 1270.9(8) \AA\/$^3$, are based upon the refinement of the XYZ-centroids of 6657 reflections above 20$\sigma$($I$) with 6.412$^\circ$ $<$ 2$\theta$ $<$ 56.35$^\circ$. Data were also corrected for absorption effects. The ratio of minimum to maximum apparent transmission was 0.724. The final anisotropic full-matrix least-squares refinement on $F^2$ with 40 variables converged at $R_1$ = 2.15\%, for the observed data and $wR_2$ = 5.16\% for all data. The goodness-of-fit was 1.139. The largest peak in the final difference electron density synthesis was 1.067 e$^-$/\AA\/$^3$ and the largest hole was -0.722 e$^-$/\AA\/$^3$ with an root-mean-square deviation of 0.146 e$^-$/\AA\/$^3$. On the basis of the final model, the calculated density was 4.28 g/cm$^3$ and F(000), 1440 e$^-$. The obtained crystal structure is shown in Fig.\ref{Fig:Fig1}(d). The stacking of perovskite slabs and rock salt layer is along [001]. Within the perovskite layers, the octahedra are connecting corners along [110] and [1$\overline{1}$0]. In the slightly distorted tetragonal $P4_2/mnm$ structure, the octahedra tilting is only present along one of the in-plane directions. The out-of-phase tilting with a tilting angle of -7.863$^\circ$ keeps alternating from one block of double-layer to the next due to the screw axis 4$_2$ around [001].
\begin{table}[ht]
\scriptsize
\begin{center}
\begin{tabular*}{\columnwidth}{ @{\extracolsep{\fill}\hspace{\tabcolsep}} lllllll @{\hspace{\tabcolsep}}}
\hline 
\hline
\multicolumn{2}{l}{Space Group} & $P4_2/mnm$ &  \multicolumn{2}{r}{Temperature} & 100 K \\ 
\hline 
\multicolumn{2}{l}{Lattice} &  $a$(\AA\/) & $b$(\AA\/) & $c$(\AA\/) \\  
\multicolumn{2}{l}{Constants} & 7.071(2) & 7.071(2) & 25.418(5) \\ 
\hline 
\multicolumn{2}{l}{Atomic Sites}  & $x/a$ & $y/b$ & $z/c$ & U(eq) \\ 
\multicolumn{2}{l}{Ba1} & 0.74627(4) & 0.74627(4) & 0.5 & 0.01153(12) \\ 
\multicolumn{2}{l}{Ba2} & 0.26045(3) & 0.73955(3) & 0.81869(2) & 0.01114(10) \\ 
\multicolumn{2}{l}{S1} & 0.2854(2) & 0.7146(2) & 0.5 & 0.0161(4) \\ 
\multicolumn{2}{l}{S2} & 0.5 & 0.5 & 0.60860(8) & 0.0127(4) \\ 
\multicolumn{2}{l}{S3} & 0.22238(15) & 0.77762(15) & 0.69788(5) & 0.0153(3) \\ 
\multicolumn{2}{l}{S4} & 0.0 & 0.0 & 0.58414(8) & 0.0113 (4) \\ 
\multicolumn{2}{l}{S5} & 0 & 0.5 & 0.59528(6) & 0.0132(3) \\ 
\multicolumn{2}{l}{Zr1} & 0.25039(5) & 0.74961(5) & 0.59974(2) & 0.00769(12) \\ 
\hline
& U$_{11}$  & U$_{22}$ & U$_{33}$ & U$_{23}$ & U$_{13}$ & U$_{12}$ \\
\hline
Ba1 & 0.01347(15) & 0.01347(15) & 0.0076(2) & 0 & 0 & 0.0014(2) \\
Ba2 & 0.01371(12) & 0.01371(12) & 0.00600(16) & -0.00022(8) & 0.00022(8) & -0.00017(13) \\
S1 & 0.0205(6) & 0.0205(6) & 0.0072(8) & 0 & 0 & 0.0030(8) \\
S2 & 0.0097(5) & 0.0097(5) & 0.0188(9) & 0 & 0 & 0.0027(7) \\
S3 & 0.0201(4) & 0.0201(4) & 0.0057(5) & -0.0001(4) & 0.0001(4) & 0.0006(6) \\
S4 & 0.0098(5) & 0.0098(5) & 0.0144(8) & 0 & 0 & 0.0024(7) \\
S5 & 0.0094(6) & 0.0094(6) & 0.0208(6) & 0 & 0 & -0.0039(5) \\
Zr1 & 0.00801(15) & 0.00801(15) & 0.0070(2) & 0.00023(11) & -0.00023(11) & 0.00014(19) \\
\hline 
\hline 
\end{tabular*} 
\caption{Space group, lattice constants, atomic coordinates, equivalent isotropic atomic displacement parameters (\AA\/$^2$), and anisotropic atomic displacement parameters (\AA\/$^2$) from X-ray crystallographic analysis for Ba$_3$Zr$_2$S$_7$.}
\label{Tab:BZS327XtalXRD}
\end{center}
\end{table}

\begin{figure}[ht]
\centering
\includegraphics[width=6.5 in]{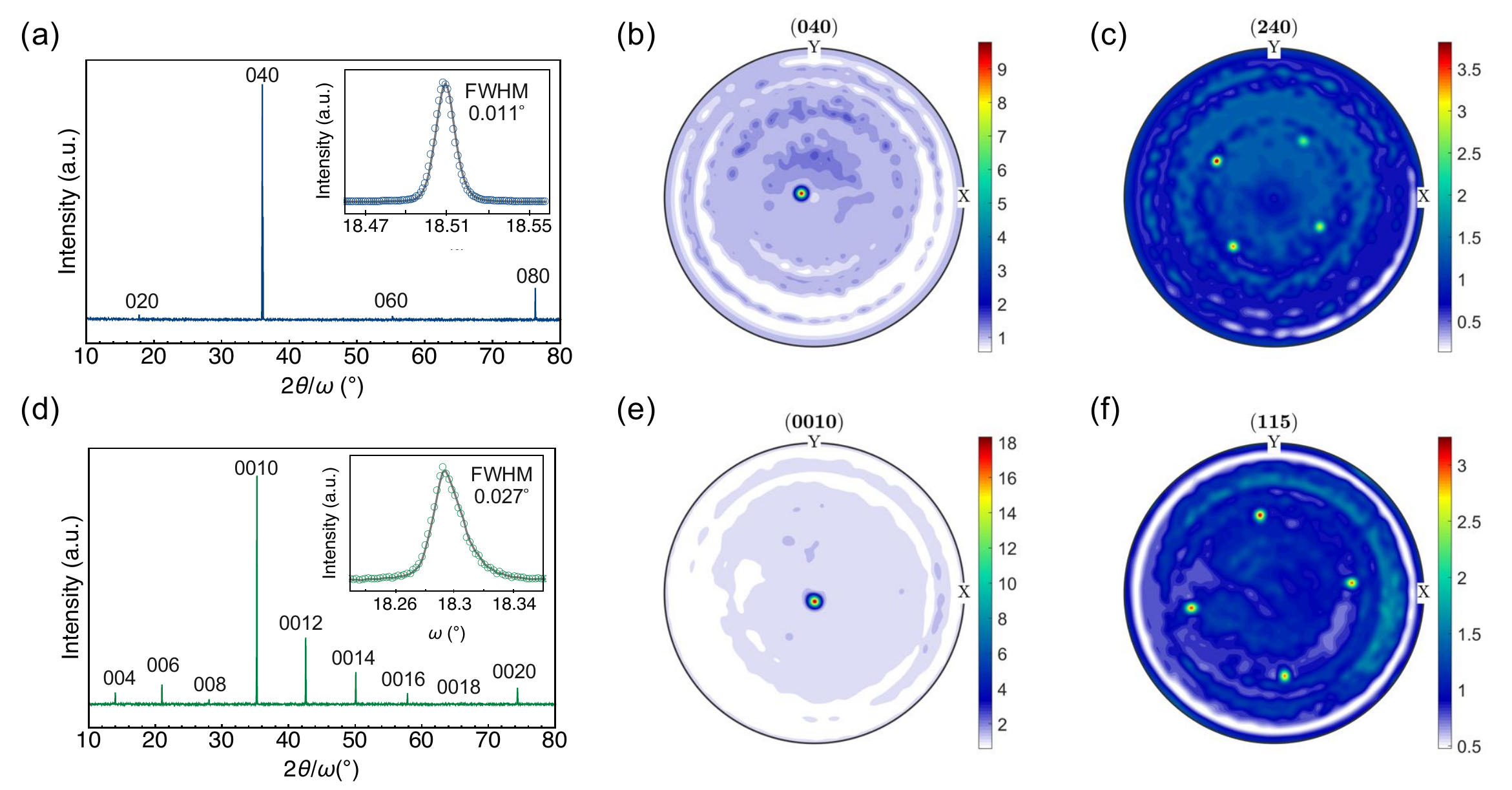}
\caption{\label{Fig:XRD} Out-of-plane XRD of individual crystal for (a) BaZrS$_3$ and (d) Ba$_3$Zr$_2$S$_7$. The insets are rocking curves of the most intense peaks. Pole figures for (b) BaZrS$_3$ {040}, (c) BaZrS$_3$ {240}, (e) Ba$_3$Zr$_2$S$_7$ {0010}, and (f) Ba$_3$Zr$_2$S$_7$ {115} plotted as orientation density function. The color bars indicate multiples of distribution function.}
\end{figure}

\subsection{Mosaicity and texture analysis}
For the bigger crystals of BaZrS$_3$ and Ba$_3$Zr$_2$S$_7$, we performed out-of-plane thin film XRD scans on isolated crystal pieces in an X-ray diffractometer with monochromatic radiation. For BaZrS$_3$, a set of narrow 0$k$0 type reflections were observed, as shown in Fig.\ref{Fig:XRD}(a). The most intense reflections have a full-width-at-half-maximum (FWHM) of less than 0.04$^\circ$. We also performed high resolution rocking curve (RC) measurements. The inset is the rocking curves of the most intense 040 reflection. Notably, we obtained a RC FWHM of 0.011$^\circ$ for BaZrS$_3$ crystal, indicating highly oriented structural quality of the single crystal. We note that cubic looking crystals could also show 101 type termination. As mentioned earlier, the difference in lattice spacing between these two directions is less than 0.3\%. out-of-plane XRD scan of Ba$_3$Zr$_2$S$_7$ showed a set of 00$l$ type peaks, as shown in Fig.\ref{Fig:XRD}(d). This proves that the terminating crystal facets with layered-like features are the (00$l$) plane. For Ba$_3$Zr$_2$S$_7$, the rocking curve width is notably larger, with a FWHM of 0.027$^\circ$ for 0010 reflection, presumably arising from higher degree of mosaicity associated with the layered nature of the crystal structure. The phase and orientation of the single crystal samples were confirmed with X-ray pole figure analysis. The {040} and {240} pole figures of a BaZrS$_3$ crystal are shown in Fig.\ref{Fig:XRD}(b), (c). Fig.\ref{Fig:XRD}(b) indicates (040) planes are aligned to the normal of the sample surface, confirming the termination facet. The spot is slightly off the center due to the small misorientation of the sample when mounted on the stage. Fig.\ref{Fig:XRD}(c) exhibits the 4 fold symmetry of the (042) poles, resulting from very similar lattice parameters for $a$ and $c$ lattice parameters. The {0010} and {115} pole figures of a Ba$_3$Zr$_2$S$_7$ crystal were shown in Fig.\ref{Fig:XRD}(e), (f). Similarly, Fig.\ref{Fig:XRD}(e) confirms (00l) termination and Fig.\ref{Fig:XRD}(f) exhibits the 4 fold symmetry of the (115) poles in the tetragonal crystal system.

\subsection{High resolution electron microscopy}
\begin{figure}[ht]
\centering
\includegraphics[width=6.5 in]{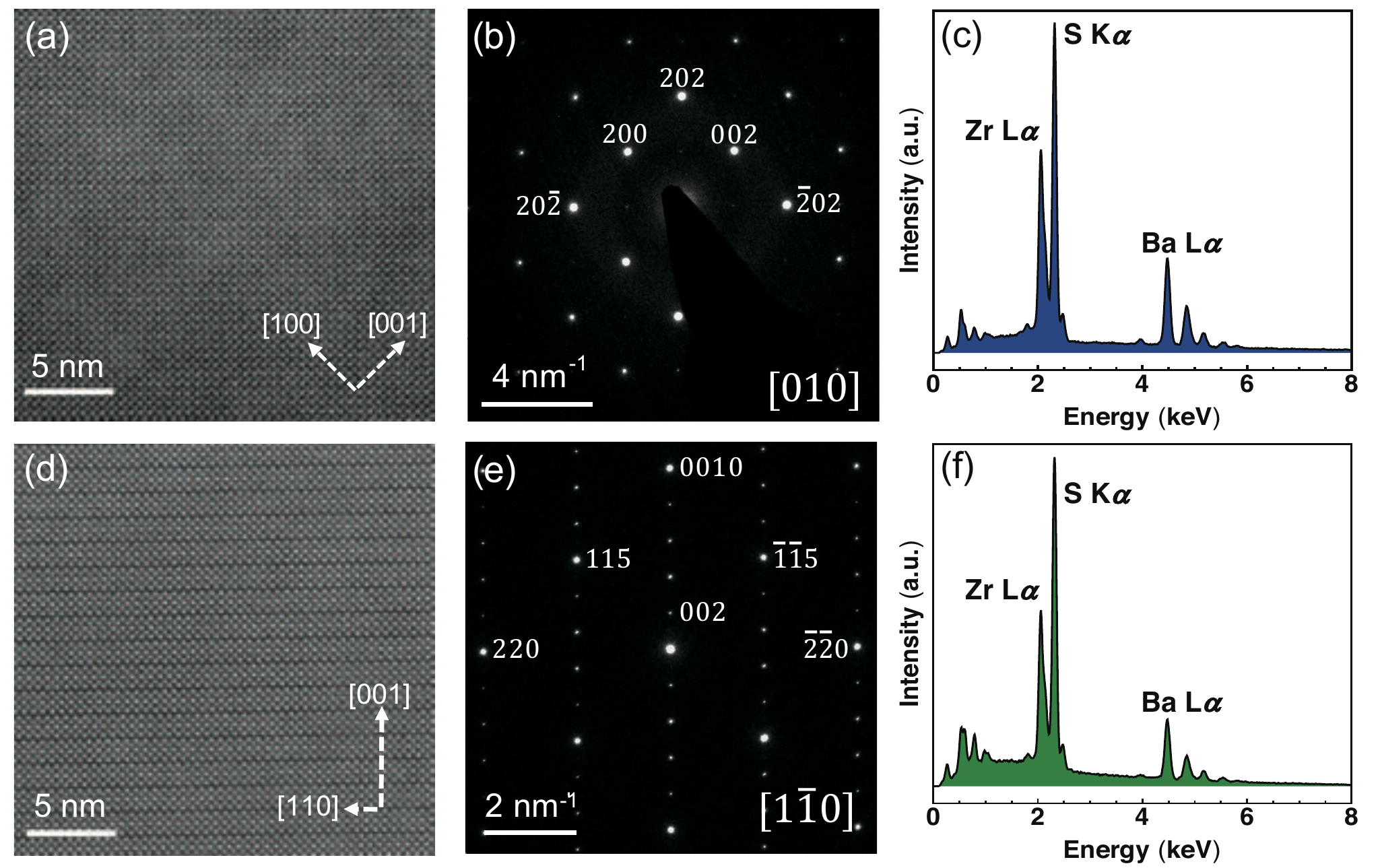}
\caption{\label{Fig:TEM} The cross-sectional STEM images of (a) BaZrS$_3$ viewed along $b$ axis and (d) Ba$_3$Zr$_2$S$_7$ viewed along (110) direction. Corresponding SAED patterns of of (b) BaZrS$_3$ and (e) Ba$_3$Zr$_2$S$_7$. EDS spectra of (c) BaZrS$_3$and (f) Ba$_3$Zr$_2$S$_7$ samples taken at 400X. The obtained Ba:Zr:S ratios are 1.06:1:3.04 and 1.56:1.1:3.45, respectively. }
\end{figure}

We further performed scanning transmission electron microscopy (STEM) studies on the crystals. The high-angle annular dark-field (HAADF) image of a BaZrS$_3$ and a Ba$_3$Zr$_2$S$_7$ crystal and corresponding selected area electron diffraction (SAED) patterns are shown in Fig.\ref{Fig:TEM}. In Fig.\ref{Fig:TEM}(a), we can see the extended 3D corner sharing octahedra network and a pseudo-cubic pattern in BaZrS$_3$. We were not able to resolve the subtle Ba displacements in STEM and the HAADF collection angle excludes scattering from the lighter S atoms. Given the small difference in spacing between the three quasi-cubic directions, (020), (101) and (10$\overline{1}$), the image and diffraction pattern can be indexed with either [010] or [101] being the zone axis. In Ba$_3$Zr$_2$S$_7$, as shown in Fig.\ref{Fig:TEM}(d), one can see the double-layer perovskite blocks stacked along [001], the blocks are offset by half a unit cell along the face diagonal of the in-plane square lattice. The diffraction pattern in Fig.\ref{Fig:TEM}(e) can be indexed with a zone axis of [1$\overline{1}$0] (or equivalently [110]). The denser diffraction spots along (001) is due to the much larger lattice constants of $c$ axis. The bright 0010 spot and weaker 004, 006, 008 spots resemble very well the intensity of the peaks we see in the out-of-plane scan in Fig.\ref{Fig:XRD}(d). Such highly symmetric SAED patterns indicate high crystallinity of the grown crystals over the selected region and agree well with the pole figure analysis. Chemical composition analysis by EDS with varying locations and magnifications on the crystals showed only expected elements in a consistent ratio, as shown in Fig.\ref{Fig:TEM}(c),(f).

\section{Conclusion}
The growth of high quality BaZrS$_3$ crystals was achieved using flux growth method with BaCl$_2$ flux. Through X-ray diffraction analysis of the crystal, space group of $Pnma$ with lattice constants of $a$ = 7.056(3)\AA\/, $b$ = 9.962(4)\AA\/, $c$ = 6.996(3)\AA\/ was extracted for BaZrS$_3$. The octahedra tilting system was identified to be $a^+b^-b^-$, with in-phase tilting of 9.0005$^\circ$ around [010] and out-of-phase tilting of -7.027$^\circ$ around [101] and [10$\overline{1}$]. STEM images and electron diffraction pattern were used to probe the local crystallinity of the crystal. High resolution X-ray rocking curve (FWHM of 0.011$^\circ$) and pole figure analysis for {040} and {240} revealed overall high crystallinity of the grown crystal. Ba$_3$Zr$_2$S$_7$ crystal was also grown using similar method by varying the starting precursor stoichiometry ratio. X-ray diffraction analysis of Ba$_3$Zr$_2$S$_7$ crystal showed a $P4_2/mnm$ structure and a rocking curve FWHM of 0.027$^\circ$. STEM image with [110] zone axis and corresponding diffraction pattern, along with pole figure analysis for {0010} and {115} were shown to further establish high crystallinity. We expect the crystal growth of perovskite chalcogenides to enable more advanced spectroscopic and transport measurements to study their optoelectronic properties.

\section{Experimental section}
\subsection{Materials synthesis}
The single crystals were grown using salt flux method in sealed quartz ampoules with BaCl$_2$ flux. Barium Sulfide powder (Sigma-Aldrich 99.9\%), Zirconium powder (STREM, 99.5\%), and Sulfur pieces (Alfa Aesar 99.999\%) were stored and handled in an Argon-filled glove box. Stoichiometric quantities of precursor powders with a total weight of 0.5 g were mixed and loaded into a 3/4 inch diameter quartz tube with 1.5 mm thickness along with around 0.5g BaCl$_2$ (Alfa Aesar 99.998\%) inside the glove box. The tube was capped with ultra-torr fittings and a bonnet needle valve to avoid exposure to the air, until it was evacuated and sealed using a blowtorch. The sealed tubes were about 12 cm in length. The tubes were heated to 1050 $^\circ$C with a ramping rate of 0.3 $^\circ$C/min, held at 1050 $^\circ$C for 40 hours and then cooled to 400 $^\circ$C with a cooling rate of 1 $^\circ$C/min, and then allowed to naturally cool down after that by shutting off the furnace. The obtained samples were washed with deionized water repeatedly to remove the excess flux, and dried with acetone and isopropyl alcohol. In some cases, residue flux after washing were removed by sonicating the crystals in isopropyl alcohol.
\subsection{Rocking curve measurement}
The thin film X-ray diffraction and rocking curve measurements were carried out in a Bruker D8 Advance X-ray diffractometer in parallel beam configuration, using a germanium (004) two-bounce monochromator for Cu K$_{\alpha 1}$  radiation. 
\subsection{Pole figure analysis}
X-ray pole figure analysis was carried out using a Bruker D8 Discover GADDS with a Co K$_{\alpha}$ source, 1/4 Eulerian cradle and Vantec-2000 area detector. The pole figure data was collected with 1$^\circ$ resolution in rotation at tilts of 25$^\circ$, 50$^\circ$, 75$^\circ$ to ensure coverage over the entire pole sphere.  The two-dimensional diffraction data was reduced with Bruker Multex area 2 software and the pole figures were analyzed in MTEX texture analysis software.\cite{bachmann:2010}

\subsection{Single crystal x-ray diffraction}
The single crystal X-ray diffraction data were collected on a Bruker SMART APEX DUO 3-circle platform diffractometer and using Mo K$_\alpha$ radiation ($\lambda$ = 0.71073 \AA\/) monochromatized by a TRIUMPH curved-crystal monochromator. The diffractometer was equipped with an APEX II CCD detector and an Oxford Cryosystems Cryostream 700 apparatus for low-temperature data collection. The crystals were mounted in a Cryo-Loop using Paratone oil. The measurement was performed at 100 K. A complete hemisphere of data was scanned on $\omega$ (0.5$^\circ$) at a detector distance of 50 mm and a resolution of 512 $\times$ 512 pixels. A total of 2520 frames were collected. The frames were integrated with the Bruker SAINT software package using a SAINT V8.38A (Bruker AXS, 2013) algorithm. The structure was solved and refined using the Bruker SHELXTL Software Package.\cite{Sheldrick:2007cya}

\subsection{Electron microscopy}
Scanning electron microscopy (SEM) images were obtained in a JEOL JEM-7001F analytical field-emission scanning electron microscope equipped with EDAX Apollo X 10 mm$^2$ EDS. The SEM images and energy dispersive X-ray spectroscopy (EDS) spectra were acquired with 15 kV accelerating voltage and a working distance of 15 mm. A thin layer of Pt was sputter coated on the crystals to mitigate the charging effects. Magnifications ranging from 100$\times$ to 10000$\times$ were used to check the consistency of chemical composition ratios across the crystals. The spectra shown were recorded at 400$\times$. We used the commercially pre-analyzed BaS as a standard, and measured its EDS side by side with the crystals to serve as calibration for Ba:S ratios. The BaS powders were sealed under argon until entering SEM chamber to minimize potential oxidation. The Ba:S ratio obtained in BaS across multiple magnifications was consistently to be 1.14:1, justifying the need for such calibration. 

The scanning transmission electron microscopy (STEM) images were acquired using an aberration-corrected FEI Titan (Thermo Fisher Scientific Inc.) transmission electron microscope operated at an accelerating voltage of 300 kV. Samples for TEM observations were prepared by focused ion beam (FIB) milling using a Ga-ion beam at an accelerating voltage of 30 kV, followed by a cleaning/polishing procedure by Ar-ion milling at 1.5 kV and 700 V to remove FIB induced residual Ga and surface amorphization in the sample.

\section{Acknowledgements}
J.R., S.N. and B.Z. acknowledge the Air Force Office of Scientific Research under award number FA9550-16-1-0335 and Army Research Office under award number W911NF-19-1-0137. S.N. acknowledges Link Foundation Energy Fellowship. M.E.M. and E.B. acknowledges support by the Air Force Office of Scientific Research under award number FA9550-15RXCOR198. E.B. acknowledges the National Science Foundation Graduate Research Fellowship under Grant No. DGE-1450681. The authors gratefully acknowledge the use of facilities at Core Center of Excellence in Nano Imaging at University of Southern California and the use of facilities and instrumentation supported by NSF through the Massachusetts Institute of Technology Materials Research Science and Engineering Center DMR-1419807.

\begin{suppinfo}
The following files are available in supplementary information.
\begin{itemize}
  \item BaZrS$_3$.cif: cif file from X-ray diffraction analysis of BaZrS$_3$ crystal
  \item Ba$_3$Zr$_2$S$_7$.cif: cif file from X-ray diffraction analysis of Ba$_3$Zr$_2$S$_7$ crystal
\end{itemize}
\end{suppinfo}
\bibliography{Manuscript.bib}

\providecommand{\latin}[1]{#1}
\makeatletter
\providecommand{\doi}
  {\begingroup\let\do\@makeother\dospecials
  \catcode`\{=1 \catcode`\}=2 \doi@aux}
\providecommand{\doi@aux}[1]{\endgroup\texttt{#1}}
\makeatother
\providecommand*\mcitethebibliography{\thebibliography}
\csname @ifundefined\endcsname{endmcitethebibliography}
  {\let\endmcitethebibliography\endthebibliography}{}
\begin{mcitethebibliography}{30}
\providecommand*\natexlab[1]{#1}
\providecommand*\mciteSetBstSublistMode[1]{}
\providecommand*\mciteSetBstMaxWidthForm[2]{}
\providecommand*\mciteBstWouldAddEndPuncttrue
  {\def\EndOfBibitem{\unskip.}}
\providecommand*\mciteBstWouldAddEndPunctfalse
  {\let\EndOfBibitem\relax}
\providecommand*\mciteSetBstMidEndSepPunct[3]{}
\providecommand*\mciteSetBstSublistLabelBeginEnd[3]{}
\providecommand*\EndOfBibitem{}
\mciteSetBstSublistMode{f}
\mciteSetBstMaxWidthForm{subitem}{(\alph{mcitesubitemcount})}
\mciteSetBstSublistLabelBeginEnd
  {\mcitemaxwidthsubitemform\space}
  {\relax}
  {\relax}

\bibitem[Sun \latin{et~al.}(2015)Sun, Agiorgousis, Zhang, and
  Zhang]{Sun:2015be}
Sun,~Y.-Y.; Agiorgousis,~M.~L.; Zhang,~P.; Zhang,~S. {Chalcogenide Perovskites
  for Photovoltaics}. \emph{Nano Letters} \textbf{2015}, \emph{15},
  581--585\relax
\mciteBstWouldAddEndPuncttrue
\mciteSetBstMidEndSepPunct{\mcitedefaultmidpunct}
{\mcitedefaultendpunct}{\mcitedefaultseppunct}\relax
\EndOfBibitem
\bibitem[Wang \latin{et~al.}(2016)Wang, Gou, and Li]{Wang:2016fh}
Wang,~H.; Gou,~G.; Li,~J. {Ruddlesden{\textendash}Popper perovskite sulfides
  A$_{3}$B$_{2}$S$_{7}$: A new family of ferroelectric photovoltaic materials
  for the visible spectrum}. \emph{Nano Energy} \textbf{2016}, \emph{22},
  507--513\relax
\mciteBstWouldAddEndPuncttrue
\mciteSetBstMidEndSepPunct{\mcitedefaultmidpunct}
{\mcitedefaultendpunct}{\mcitedefaultseppunct}\relax
\EndOfBibitem
\bibitem[Kuhar \latin{et~al.}(2017)Kuhar, Crovetto, Pandey, Thygesen, Seger,
  Vesborg, Hansen, Chorkendorff, and Jacobsen]{Kuhar:2017gw}
Kuhar,~K.; Crovetto,~A.; Pandey,~M.; Thygesen,~K.~S.; Seger,~B.; Vesborg,~P.
  C.~K.; Hansen,~O.; Chorkendorff,~I.; Jacobsen,~K.~W. {Sulfide perovskites for
  solar energy conversion applications: computational screening and synthesis
  of the selected compound LaYS$_{3}$}. \emph{Energy {\&} Environmental
  Science} \textbf{2017}, \emph{10}, 2579--2593\relax
\mciteBstWouldAddEndPuncttrue
\mciteSetBstMidEndSepPunct{\mcitedefaultmidpunct}
{\mcitedefaultendpunct}{\mcitedefaultseppunct}\relax
\EndOfBibitem
\bibitem[Ju \latin{et~al.}(2017)Ju, Dai, Ma, and Zeng]{Ju:2017iw}
Ju,~M.-G.; Dai,~J.; Ma,~L.; Zeng,~X.~C. {Perovskite chalcogenides with optimal
  bandgap and desired optical absorption for photovoltaic devices}.
  \emph{Advanced Energy Materials} \textbf{2017}, \emph{48}, 1700216\relax
\mciteBstWouldAddEndPuncttrue
\mciteSetBstMidEndSepPunct{\mcitedefaultmidpunct}
{\mcitedefaultendpunct}{\mcitedefaultseppunct}\relax
\EndOfBibitem
\bibitem[Niu \latin{et~al.}(2018)Niu, Joe, Zhao, Zhou, Orvis, Huyan, Salman,
  Mahalingam, Urwin, Wu, Liu, Tiwald, Cronin, Howe, Mecklenburg, Haiges, Singh,
  Wang, Kats, and Ravichandran]{Niu:2018doa}
Niu,~S.; Joe,~G.; Zhao,~H.; Zhou,~Y.; Orvis,~T.; Huyan,~H.; Salman,~J.;
  Mahalingam,~K.; Urwin,~B.; Wu,~J.; Liu,~Y.; Tiwald,~T.~E.; Cronin,~S.~B.;
  Howe,~B.~M.; Mecklenburg,~M.; Haiges,~R.; Singh,~D.~J.; Wang,~H.;
  Kats,~M.~A.; Ravichandran,~J. {Giant optical anisotropy in a
  quasi-one-dimensional crystal}. \emph{Nature Photonics} \textbf{2018},
  \emph{12}, 392--396\relax
\mciteBstWouldAddEndPuncttrue
\mciteSetBstMidEndSepPunct{\mcitedefaultmidpunct}
{\mcitedefaultendpunct}{\mcitedefaultseppunct}\relax
\EndOfBibitem
\bibitem[Niu \latin{et~al.}(2018)Niu, Sarkar, Williams, Zhou, Li, Bianco,
  Huyan, Cronin, McConney, Haiges, Jaramillo, Singh, Tisdale, Kapadia, and
  Ravichandran]{Niu:2018hu}
Niu,~S.; Sarkar,~D.; Williams,~K.; Zhou,~Y.; Li,~Y.; Bianco,~E.; Huyan,~H.;
  Cronin,~S.~B.; McConney,~M.~E.; Haiges,~R.; Jaramillo,~R.; Singh,~D.~J.;
  Tisdale,~W.~A.; Kapadia,~R.; Ravichandran,~J. {Optimal Bandgap in a 2D
  Ruddlesden{\textendash}Popper Perovskite Chalcogenide for Single-Junction
  Solar Cells}. \emph{Chemistry of Materials} \textbf{2018}, \emph{30},
  4882--4886\relax
\mciteBstWouldAddEndPuncttrue
\mciteSetBstMidEndSepPunct{\mcitedefaultmidpunct}
{\mcitedefaultendpunct}{\mcitedefaultseppunct}\relax
\EndOfBibitem
\bibitem[Filippone \latin{et~al.}(2018)Filippone, Sun, and
  Jaramillo]{Filippone:2018cr}
Filippone,~S.~A.; Sun,~Y.-Y.; Jaramillo,~R. {Determination of
  adsorption-controlled growth windows of chalcogenide perovskites }. \emph{MRS
  Communications} \textbf{2018}, \emph{8}, 145--151\relax
\mciteBstWouldAddEndPuncttrue
\mciteSetBstMidEndSepPunct{\mcitedefaultmidpunct}
{\mcitedefaultendpunct}{\mcitedefaultseppunct}\relax
\EndOfBibitem
\bibitem[Niu \latin{et~al.}(2018)Niu, Zhao, Zhou, Huyan, Zhao, Wu, Cronin,
  Wang, and Ravichandran]{Niu:2018cx}
Niu,~S.; Zhao,~H.; Zhou,~Y.; Huyan,~H.; Zhao,~B.; Wu,~J.; Cronin,~S.~B.;
  Wang,~H.; Ravichandran,~J. {Mid-wave and Long-Wave Infrared Linear Dichroism
  in a Hexagonal Perovskite Chalcogenide}. \emph{Chemistry of Materials}
  \textbf{2018}, \emph{30}, 4897--4901\relax
\mciteBstWouldAddEndPuncttrue
\mciteSetBstMidEndSepPunct{\mcitedefaultmidpunct}
{\mcitedefaultendpunct}{\mcitedefaultseppunct}\relax
\EndOfBibitem
\bibitem[Hanzawa \latin{et~al.}(2019)Hanzawa, Iimura, Hiramatsu, and
  Hosono]{Hanzawa:2019cf}
Hanzawa,~K.; Iimura,~S.; Hiramatsu,~H.; Hosono,~H. {Material Design of
  Green-Light-Emitting Semiconductors: Perovskite-Type Sulfide SrHfS$_3$}.
  \emph{Journal of the American Chemical Society} \textbf{2019},
  jacs.8b13622\relax
\mciteBstWouldAddEndPuncttrue
\mciteSetBstMidEndSepPunct{\mcitedefaultmidpunct}
{\mcitedefaultendpunct}{\mcitedefaultseppunct}\relax
\EndOfBibitem
\bibitem[Swarnkar \latin{et~al.}(2019)Swarnkar, Mir, Chakraborty,
  Jagadeeswararao, Sheikh, and Nag]{Swarnkar:2019ia}
Swarnkar,~A.; Mir,~W.~J.; Chakraborty,~R.; Jagadeeswararao,~M.; Sheikh,~T.;
  Nag,~A. {Are Chalcogenide Perovskites an Emerging Class of Semiconductors for
  Optoelectronic Properties and Solar Cell?} \emph{Chemistry of Materials}
  \textbf{2019}, \relax
\mciteBstWouldAddEndPunctfalse
\mciteSetBstMidEndSepPunct{\mcitedefaultmidpunct}
{}{\mcitedefaultseppunct}\relax
\EndOfBibitem
\bibitem[Bennett \latin{et~al.}(2009)Bennett, Grinberg, and
  Rappe]{Bennett:2009hn}
Bennett,~J.~W.; Grinberg,~I.; Rappe,~A.~M. {Effect of substituting of S for O:
  the sulfide perovskite BaZrS$_{3}$ investigated with density functional
  theory}. \emph{Physical Review B} \textbf{2009}, \emph{79}, 235115\relax
\mciteBstWouldAddEndPuncttrue
\mciteSetBstMidEndSepPunct{\mcitedefaultmidpunct}
{\mcitedefaultendpunct}{\mcitedefaultseppunct}\relax
\EndOfBibitem
\bibitem[Hahn and Mutschke(1957)Hahn, and Mutschke]{Hahn:1956wo}
Hahn,~H.; Mutschke,~U. {Untersuchungen {\"u}ber tern{\"a}re Chalkogenide. XI.
  Versuche zur Darstellung von Thioperowskiten}. \emph{Zeitschrift Fur
  Anorganische Und Allgemeine Chemie} \textbf{1957}, \emph{288}, 269--278\relax
\mciteBstWouldAddEndPuncttrue
\mciteSetBstMidEndSepPunct{\mcitedefaultmidpunct}
{\mcitedefaultendpunct}{\mcitedefaultseppunct}\relax
\EndOfBibitem
\bibitem[Clearfield(1963)]{CLEARFIELD:1963ip}
Clearfield,~A. {The synthesis and crystal structures of some alkaline earth
  titanium and zirconium sulfides}. \emph{Acta Crystallographica}
  \textbf{1963}, \emph{16}, 135--142\relax
\mciteBstWouldAddEndPuncttrue
\mciteSetBstMidEndSepPunct{\mcitedefaultmidpunct}
{\mcitedefaultendpunct}{\mcitedefaultseppunct}\relax
\EndOfBibitem
\bibitem[Lelieveld and Ijdo(1980)Lelieveld, and Ijdo]{LELIEVELD:1980wp}
Lelieveld,~R.; Ijdo,~D. J.~W. {Sulphides with the GdFeO$_{3}$ structure}.
  \emph{Acta Crystallographica Section B Structural Crystallography and Crystal
  Chemistry} \textbf{1980}, \emph{36}, 2223--2226\relax
\mciteBstWouldAddEndPuncttrue
\mciteSetBstMidEndSepPunct{\mcitedefaultmidpunct}
{\mcitedefaultendpunct}{\mcitedefaultseppunct}\relax
\EndOfBibitem
\bibitem[Huster(1980)]{Huster:1980fn}
Huster,~J. {Die Kristallstruktur von BaTiS$_{3}$}. \emph{Zeitschrift f{\"u}r
  Naturforschung B} \textbf{1980}, \emph{35}, 775\relax
\mciteBstWouldAddEndPuncttrue
\mciteSetBstMidEndSepPunct{\mcitedefaultmidpunct}
{\mcitedefaultendpunct}{\mcitedefaultseppunct}\relax
\EndOfBibitem
\bibitem[Lee \latin{et~al.}(2005)Lee, Kleinke, and Kleinke]{Lee:2005et}
Lee,~C.-S.; Kleinke,~K.~M.; Kleinke,~H. {Synthesis, structure, and electronic
  and physical properties of the two SrZrS$_{3}$ modifications}. \emph{Solid
  State Sciences} \textbf{2005}, \emph{7}, 1049--1054\relax
\mciteBstWouldAddEndPuncttrue
\mciteSetBstMidEndSepPunct{\mcitedefaultmidpunct}
{\mcitedefaultendpunct}{\mcitedefaultseppunct}\relax
\EndOfBibitem
\bibitem[Okai \latin{et~al.}(1988)Okai, Takahashi, Saeki, and
  Yoshimoto]{BinOkai:1988ut}
Okai,~B.; Takahashi,~K.; Saeki,~M.; Yoshimoto,~J. {Preparation and crystal
  structures of some complex sulphides at high pressures}. \emph{Materials
  Research Bulletin} \textbf{1988}, \emph{23}, 1575--1584\relax
\mciteBstWouldAddEndPuncttrue
\mciteSetBstMidEndSepPunct{\mcitedefaultmidpunct}
{\mcitedefaultendpunct}{\mcitedefaultseppunct}\relax
\EndOfBibitem
\bibitem[Niu \latin{et~al.}(2017)Niu, Huyan, Liu, Yeung, Ye, Blankemeier,
  Orvis, Sarkar, Singh, Kapadia, and Ravichandran]{niu:2016ce}
Niu,~S.; Huyan,~H.; Liu,~Y.; Yeung,~M.; Ye,~K.; Blankemeier,~L.; Orvis,~T.;
  Sarkar,~D.; Singh,~D.~J.; Kapadia,~R.; Ravichandran,~J. {Bandgap Control via
  Structural and Chemical Tuning of Transition Metal Perovskite Chalcogenides}.
  \emph{Advanced Materials} \textbf{2017}, \emph{29}, 1604733\relax
\mciteBstWouldAddEndPuncttrue
\mciteSetBstMidEndSepPunct{\mcitedefaultmidpunct}
{\mcitedefaultendpunct}{\mcitedefaultseppunct}\relax
\EndOfBibitem
\bibitem[Hung \latin{et~al.}(1997)Hung, Fettinger, and Eichhorn]{Hung:1997gg}
Hung,~Y.~C.; Fettinger,~J.~C.; Eichhorn,~B.~W. {Ba$_{3}$Zr$_{2}$S$_{7}$, the
  low-temperature polymorph}. \emph{Acta Crystallographica Section C Crystal
  Structure Communications} \textbf{1997}, \emph{53}, 827--829\relax
\mciteBstWouldAddEndPuncttrue
\mciteSetBstMidEndSepPunct{\mcitedefaultmidpunct}
{\mcitedefaultendpunct}{\mcitedefaultseppunct}\relax
\EndOfBibitem
\bibitem[Chen \latin{et~al.}(1994)Chen, Eichhorn, and Wong-Ng]{Chen:1994ba}
Chen,~B.~H.; Eichhorn,~B.; Wong-Ng,~W. {Structural reinvestigation of
  Ba$_{3}$Zr$_{2}$S$_{7}$ by single-crystal X-ray diffraction}. \emph{Acta
  Crystallographica Section C Crystal Structure Communications} \textbf{1994},
  \emph{50}, 161--164\relax
\mciteBstWouldAddEndPuncttrue
\mciteSetBstMidEndSepPunct{\mcitedefaultmidpunct}
{\mcitedefaultendpunct}{\mcitedefaultseppunct}\relax
\EndOfBibitem
\bibitem[Chen \latin{et~al.}(1993)Chen, Wong-Ng, and Eichhorn]{Chen:1993cy}
Chen,~B.-H.; Wong-Ng,~W.; Eichhorn,~B.~W. {Preparation of New
  Ba$_{4}$M$_{3}$S$_{10}$ Phases (M = Zr, Hf) and Single Crystal Structure
  Determination of Ba$_{4}$Zr$_{3}$S$_{10}$}. \emph{Journal of Solid State
  Chemistry} \textbf{1993}, \emph{103}, 75--80\relax
\mciteBstWouldAddEndPuncttrue
\mciteSetBstMidEndSepPunct{\mcitedefaultmidpunct}
{\mcitedefaultendpunct}{\mcitedefaultseppunct}\relax
\EndOfBibitem
\bibitem[Meng \latin{et~al.}(2016)Meng, Saparov, Hong, Wang, Mitzi, and
  Yan]{Meng:2016dv}
Meng,~W.; Saparov,~B.; Hong,~F.; Wang,~J.; Mitzi,~D.~B.; Yan,~Y. {Alloying and
  Defect Control within Chalcogenide Perovskites for Optimized Photovoltaic
  Application}. \emph{Chemistry of Materials} \textbf{2016}, \emph{28},
  821--829\relax
\mciteBstWouldAddEndPuncttrue
\mciteSetBstMidEndSepPunct{\mcitedefaultmidpunct}
{\mcitedefaultendpunct}{\mcitedefaultseppunct}\relax
\EndOfBibitem
\bibitem[Perera \latin{et~al.}(2016)Perera, Hui, Zhao, Xue, Sun, Deng, Gross,
  Milleville, Xu, Watson, Weinstein, Sun, Zhang, and Zeng]{Perera:2016gf}
Perera,~S.; Hui,~H.; Zhao,~C.; Xue,~H.; Sun,~F.; Deng,~C.; Gross,~N.;
  Milleville,~C.; Xu,~X.; Watson,~D.~F.; Weinstein,~B.; Sun,~Y.-Y.; Zhang,~S.;
  Zeng,~H. {Chalcogenide perovskites {\textendash} an emerging class of ionic
  semiconductors}. \emph{Nano Energy} \textbf{2016}, \emph{22}, 129--135\relax
\mciteBstWouldAddEndPuncttrue
\mciteSetBstMidEndSepPunct{\mcitedefaultmidpunct}
{\mcitedefaultendpunct}{\mcitedefaultseppunct}\relax
\EndOfBibitem
\bibitem[Gross \latin{et~al.}(2017)Gross, Sun, Perera, Hui, Wei, Zhang, Zeng,
  and Weinstein]{Gross:2017bo}
Gross,~N.; Sun,~Y.-Y.; Perera,~S.; Hui,~H.; Wei,~X.; Zhang,~S.; Zeng,~H.;
  Weinstein,~B.~A. {Stability and Band-Gap Tuning of the Chalcogenide
  Perovskite BaZrS$_{3}$ in Raman and Optical Investigations at High
  Pressures}. \emph{Physical Review Applied} \textbf{2017}, \emph{8},
  044014\relax
\mciteBstWouldAddEndPuncttrue
\mciteSetBstMidEndSepPunct{\mcitedefaultmidpunct}
{\mcitedefaultendpunct}{\mcitedefaultseppunct}\relax
\EndOfBibitem
\bibitem[Niu \latin{et~al.}(2018)Niu, Milam-Guerrero, Zhou, Ye, Zhao, Melot,
  and Ravichandran]{Niu:2018hr}
Niu,~S.; Milam-Guerrero,~J.; Zhou,~Y.; Ye,~K.; Zhao,~B.; Melot,~B.~C.;
  Ravichandran,~J. {Thermal stability study of transition metal perovskite
  sulfides}. \emph{Journal of Materials Research} \textbf{2018}, \emph{33},
  4135--4143\relax
\mciteBstWouldAddEndPuncttrue
\mciteSetBstMidEndSepPunct{\mcitedefaultmidpunct}
{\mcitedefaultendpunct}{\mcitedefaultseppunct}\relax
\EndOfBibitem
\bibitem[Woodward(1997)]{Woodward:1997ci}
Woodward,~P.~M. {Octahedral Tilting in Perovskites. I. Geometrical
  Considerations}. \emph{Acta Cryst (1997). B53, 32-43
  [doi:10.1107/S0108768196010713]} \textbf{1997}, 1--12\relax
\mciteBstWouldAddEndPuncttrue
\mciteSetBstMidEndSepPunct{\mcitedefaultmidpunct}
{\mcitedefaultendpunct}{\mcitedefaultseppunct}\relax
\EndOfBibitem
\bibitem[Saeki \latin{et~al.}(1991)Saeki, Yajima, and Onoda]{Saeki:1991fa}
Saeki,~M.; Yajima,~Y.; Onoda,~M. {Preparation and crystal structures of new
  barium zirconium sulfides, Ba$_{2}$ZrS$_{4}$ and Ba$_{3}$Zr$_{2}$S$_{7}$}.
  \emph{Journal of Solid State Chemistry} \textbf{1991}, \emph{92},
  286--294\relax
\mciteBstWouldAddEndPuncttrue
\mciteSetBstMidEndSepPunct{\mcitedefaultmidpunct}
{\mcitedefaultendpunct}{\mcitedefaultseppunct}\relax
\EndOfBibitem
\bibitem[Bachmann \latin{et~al.}(2010)Bachmann, Hielscher, and
  Schaeben]{bachmann:2010}
Bachmann,~F.; Hielscher,~R.; Schaeben,~H. Texture Analysis with MTEX $–$ Free
  and Open Source Software Toolbox. Texture and Anisotropy of Polycrystals III.
  2010; pp 63--68\relax
\mciteBstWouldAddEndPuncttrue
\mciteSetBstMidEndSepPunct{\mcitedefaultmidpunct}
{\mcitedefaultendpunct}{\mcitedefaultseppunct}\relax
\EndOfBibitem
\bibitem[Sheldrick(2008)]{Sheldrick:2007cya}
Sheldrick,~G.~M. {A short history of SHELX}. \emph{Acta Crystallographica
  Section A} \textbf{2008}, \emph{64}, 112--122\relax
\mciteBstWouldAddEndPuncttrue
\mciteSetBstMidEndSepPunct{\mcitedefaultmidpunct}
{\mcitedefaultendpunct}{\mcitedefaultseppunct}\relax
\EndOfBibitem
\end{mcitethebibliography}
\end{document}